\documentclass[fleqn,10pt]{wlscirep}
\usepackage[utf8]{inputenc}
\usepackage[T1]{fontenc}

\usepackage[utf8]{inputenc}
\usepackage[english]{babel}
\usepackage[T1]{fontenc}
\usepackage{amsmath}
\usepackage{hyperref}

\usepackage{epsfig}
\usepackage{subfigure}
\usepackage{graphicx}
\usepackage{dcolumn}
\usepackage{stmaryrd}
\usepackage{mathrsfs}
\usepackage{pifont}
\usepackage{amsthm}
\usepackage{amssymb}
\usepackage{mathtools}
\usepackage{bm}
\usepackage{latexsym}
\usepackage{color}

\usepackage{tikz}
\usepackage{lipsum}

\def\oe#1{{ Opt.\ Express} {\bf#1}}

\def\pra#1{{ Phys.\ Rev. A\/} {\bf#1}}

\def\prl#1{{ Phys.\ Rev.\ Lett.} {\bf#1}}

\title{Quantum search algorithm on weighted databases}

\author[1,*]{Yifan Sun}
\affil[1]{Beijing Key Laboratory of Nanophotonics \& Ultrafine Optoelectronic Systems, School of Physics, Beijing Institute of Technology, 100081, Beijing, China}
\author[2,3,4,*]{Lian-Ao Wu}
\affil[2]{Department of Physics, The Basque Country University (EHU/UPV), PO Box 644, 48080 Bilbao, Spain}
\affil[3]{Ikerbasque, Basque Foundation for Science, 48011 Bilbao, Spain }
\affil[4]{EHU Quantum Center, University of the Basque Country UPV/EHU, Leioa, 48940 Biscay, Spain}
\affil[*]{yfsun@bit.edu.cn;lianao.wu@ehu.es}

\begin{abstract}
The Grover algorithm is a crucial solution for addressing unstructured search problems and has emerged as an essential quantum subroutine in various complex algorithms. By using a different approach with previous studies, this research extensively investigates Grover's search methodology within non-uniformly distributed databases, a scenario frequently encountered in practical applications. Our analysis reveals that the behavior of the Grover evolution differs significantly when applied to non-uniform databases compared to uniform or `unstructured databases.' Based on the property of differential equation, it is observed that the search process facilitated by this evolution does not consistently result in a speed-up, and we have identified specific criteria for such situations. Furthermore, we have extended this investigation to databases characterized by coherent states, confirming the speed-up achieved through Grover evolution via rigorous numerical verification. In conclusion, our study provides an enhancement to the original Grover algorithm, offering insights to optimize implementation strategies and broaden its range of applications.
\end{abstract}
\begin{document}

\flushbottom
\maketitle
%
%
\thispagestyle{empty}

\section*{Introduction}
The Grover algorithm, conceived by L. K. Grover in 1997~\cite{Grover1997}, marked a significant advancement in the field of quantum computing \cite{Feynman1982}, particularly in addressing the challenge of query complexity. In the classical paradigm, searching an unstructured database typically necessitates $n$ steps, where $n$ is the size of the database. Grover's groundbreaking algorithm, however, revolutionizes this approach by reducing the required steps to merely $\sqrt{n}$. This quantum search algorithm has emerged as a cornerstone in the development of quantum computational routines, celebrated for its ability to significantly amplify the amplitude of the quantum state that encodes the desired information. 
The versatility and applicative potential of the Grover algorithm have been demonstrated across a spectrum of challenging problems. For instance, it has provided innovative solutions to the satisfiability problem~\cite{Karp1972}, as well as in the burgeoning field of quantum machine learning \cite{Weibe2015}. Further applications include tackling constrained polynomial binary optimization \cite{Gilliam2021} and enhancing quantum amplitude estimation techniques \cite{Brassard2002}, showcasing a clear computational superiority over traditional methods. 
Recent explorations have extended the utility of the Grover algorithm \cite{Ren2020} to the domain of adiabatic quantum computing \cite{Wu2002,Pyshkin2016,Sun2020}, underscoring its adaptability and relevance in the rapidly evolving landscape of quantum research. This paper specifically delves into the algorithm's seminal application in database searching, highlighting its transformative impact and ongoing significance in the quest for efficient quantum computing solutions. Through this focus, we aim to illuminate the enduring value and broad applicability of Grover's algorithm \cite{Grover1997,Tezuka2022,Reitzner2019}, from its initial proposal to its current and potential future contributions to quantum computing and beyond.

The search problem unfolds as follows: within a given database, each element is distinctly indexed. When the database is of finite size, locating a specific element necessitates iterative queries to its index. Typically, the query count scales with the database size. Grover's seminal work explored this quandary in the realm of quantum computing. Through specific evolution operators, the amplitude of the basis state housing the target data can be boosted to unity. The steps required for such enhancement scale proportionally to the square root of the database size, ensuring a guaranteed quadratic acceleration. This concept has been integrated into numerous platforms \cite{Brickman2005, Figgatt2017, Vemula2205}, with additional advancements showcased in recent proposals \cite{Gustiani2021, Pan2021, Ji2022, Ji2022-2}.


The initial discourse on the search dilemma predominantly centers on managing unstructured databases, following a conventional approach in theoretical computational discussions that remains detached from specific physical contexts. However, as we transcend the limits imposed by current computing platforms and strive for advancements, particularly in the evolution of novel computing architectures, data encoding states may not uniformly distribute. Thus, delving into the potential enhancements of the Grover search algorithm in such scenarios presents an opportunity to broaden its utility and applicability significantly.


Moving forward, we conduct an analysis of the aforementioned issue. The database under scrutiny is comprehensive, characterized by distributions spanning various types. Similar topics have been discussed before \cite{Biham1999,Biham2000}, in which a general form of the Grover evolution is given. 
Different from the previous method, we analysis the necessary steps for executing the search operation by using the asymptotic differential equation of the algorithm. 
By using the property of differential equation, we methodically identify the prerequisites for achieving acceleration through Grover evolution. Subsequently, we showcase two illustrative examples to elucidate our observations: the first example validates the harmony between our theory and Grover's established results, while the second one exemplifies that employing the Grover search on a database governed by coherent state probabilities leads to acceleration compared to conventional methods. It indicates our algorithms can be carried out in non-universal quantum computation like linear optical system used in implementing Boson sampling. The initial state can be prepared as weak coherent states depending on cutting of $N$. This is followed by an elaborate exposition of our overarching methodology.


\subsection*{Grover Search on weighted databases}\label{sec:GSonWD}
\begin{figure}[htbp]
\centering
\includegraphics[width=2.8in]{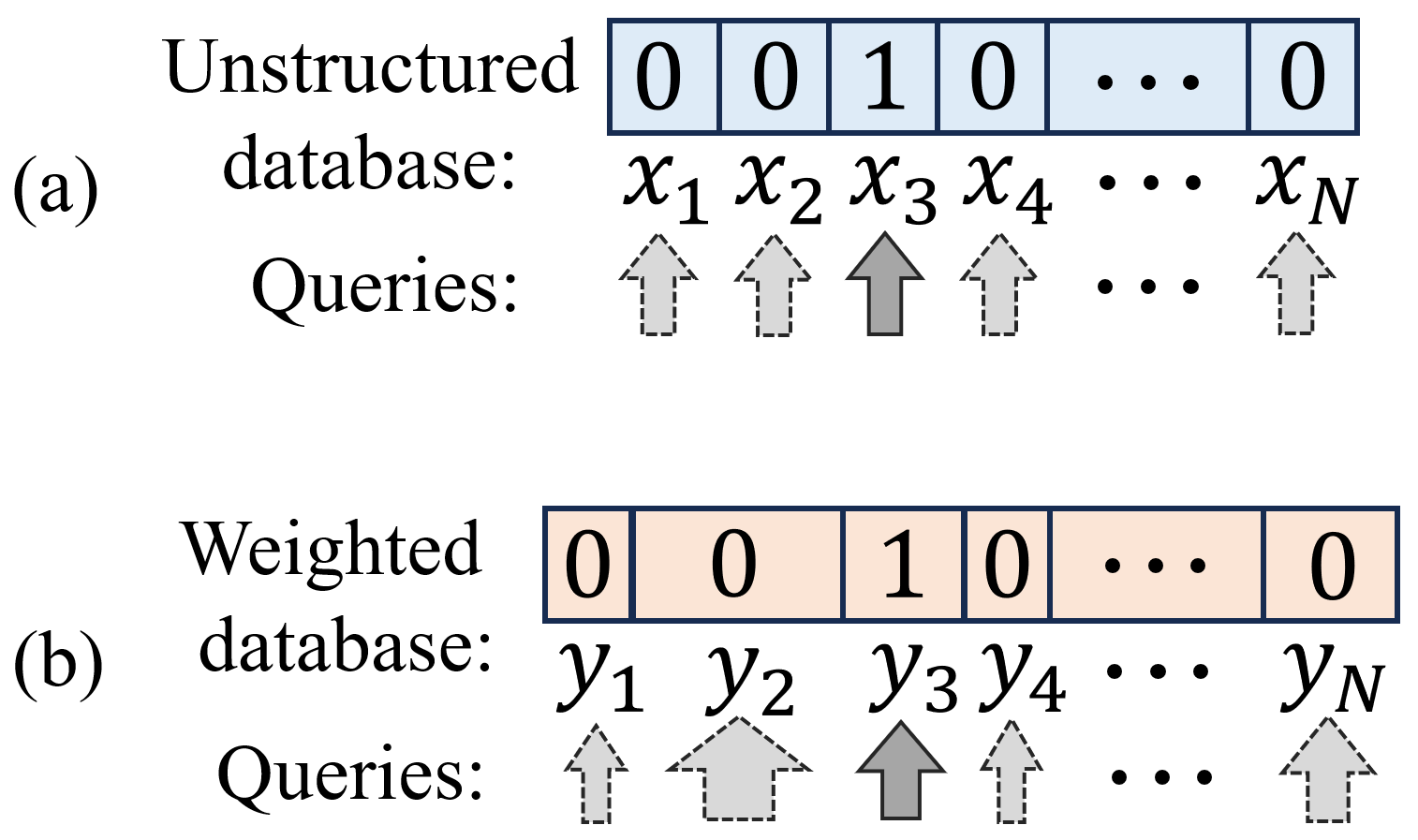}
\caption{An illustration of the search problem on 
 unstructured database (a) and weighted database(b). ``0'' and ``1'' mark the ordinary and the target data sample respectively. The widths of the squares of the samples represent the proportions.}\label{Fig:Database}
\end{figure}

Consider a database $\{x_1,\dots,x_M\}$, with integer $M$. An arbitrary element $x_n$ in the database is a real number, which represents a certain characteristic of objects. In the original version of search problems, all $x_m$s are distinct to each other. A example of the problem could be as follows. One could consider the above database as a collection of the length data of many pencils, and $x_n$ represents the length of the $n$th pencil. If the length of pencils are measured precisely enough, there will be no pencils with the same lengths. Hence, the search for the series number of the pencil with a particular length in the database could be the described by the search problem of this kind. However, in practice, some characters of the objects do not need to be handled at an extremely high precision. Going back to the pencil example, if one re-perform the statics and categorize the lengths of the pencils into several length intervals, there would be more than one pencil in one interval. Furthermore, if the pencils whose lengths fall into one length interval are numbered identically, a natural target under this case could be finding the series number of a demanded length interval. This goes into the problem of what we called the {\it weighted database search}. Formally, 
suppose that there are $N$ distinguished types of elements in a database, denoted by $y_1,\dots,y_N$. Therefore, the database 
can be given by $\{(y_1,p_1),\dots,(y_N,p_N)\}$, where $p_1,\dots,p_M$ represent the proportions of distinct characteristics $y_1,\dots,y_N$ in the total database. An illustration of the search problem on the databases is given by Fig. \ref{Fig:Database}.

To search for a certain characteristics in $\{(y_1,p_1)$, $\dots,(y_N,p_N)\}$ by using classical algorithms, the required number of steps $s$ is proportional to the reciprocal number of its proportion. Therefore, in general, $s$ satisfies 
\begin{equation}
	\min\limits_{j=1,\dots,N}\left\{\frac{1}{p_j}\right\}\leqslant s\leqslant \max\limits_{j=1,\dots,N}\left\{\frac{1}{p_j}\right\}.
\end{equation} 
To perform the same task by using Grover evolution \cite{Nielsen2000}, one can consider the following state 
\begin{equation}\label{eq:state}
|D\rangle=\sum^{N}_{n=0}P(n)|n\rangle.
\end{equation}
$\{|1\rangle,|2\rangle,\cdots,|N\rangle\}$ is a set of orthonormal basis and $|D\rangle$ is a superposition of them. $P(n)$ is the complex amplitude of the basis state $|n\rangle$, yielding that $|P(n)|^2=p_n$, and $\sum_{n=0}^N|P(n)|^2=1$. Notice that Eq. (\ref{eq:state}) can represent a state whose amplitudes of each basis state can be arbitrary. In the usual consideration of quantum computing, it is not always easy to prepare a state according to the distribution of given database. It is known that, for instance, the quantum state for encoding unstructured database, as considered by Grover, can be efficiently prepared by Hadamard gates on qubits. Also, the database distributed like a coherent state, which is an exponential distribution modified by several factors, can be prepared by displacements on harmonic oscillators in a continues-variable quantum computing setup. We will discuss the two cases later. In a general sense, a quantum computer could be any of quantum systems that can be well controlled. Hence, the initial state of the system, such as its ground states, might be not distributed uniformly in its natural basis. This could also be applied for searching the database with the same or close distribution, as indicated by Eq. (\ref{eq:state}). Because our aim in this work is discussing the Grover search based on Eq. (\ref{eq:state}) in a general sense, we do not go further into the preparation strategy of states with other distribution. We would like to discuss them broadly in our future investigation.

According to the idea of Grover search, the amplitude of the target state can be amplified by repetitive evolution so that the search can be completed by only one step. Then, the total step number of performing the search equals to the repeat number of the evolution operators. Suppose that the target state is $|k\rangle$, the basic two operators for evolution are defined by
\begin{equation}
	U_{D}=2|D\rangle\langle D|-1,\ U_k=1-2|k\rangle\langle k|.
\end{equation}
The amplification operator required by Grover search is defined by $G\coloneqq U_{D}U_k$. Suppose that after performing $G$ for $t$ times on $|D\rangle$, the whole state evolves to $|k\rangle$. Then, the step number for searching $|k\rangle$ is given by $t$. 

The next key problem is to compare the step numbers of the two methods, validating whether a speed-up exists. For such purpose, we analyze the above evolution under $G$ as follows. Applying $G$ once to the state $|D\rangle$, one has
\begin{equation}
    G|D\rangle=(1-4|P(k)|^2)|D\rangle+2P(k)|k\rangle.
\end{equation}
Furthermore, if $G$ is applied for $r$ times, a recurrence relation can be obtained,
\begin{equation}
	G^r|D\rangle=a_r|D\rangle+b_r|k\rangle,
\end{equation}
where
\begin{equation}\label{Dif_Eq}
\begin{split}
    a_r=[1&-4|P(k)|^2]a_{r-1}-2P^{*}(k)b_{r-1}, \\
        b_r&=b_{r-1}+2P(k)a_{r-1}.
\end{split}
\end{equation}
For sufficient large $r$, the amplification leads to that $a_r\to 0$ and $b_r\to 1$. This asymptotic behavior can be seen by approximating $a_r$ with a continuous function $f_a(x)$ with real variable $x$, such that $f_a(r)=a_r$. Apply the approximation 
$a_r-a_{r-1}\sim \partial f_a/\partial x$, and the same for $b_r$. Thereafter, two partial differential equations can be obtained, shown by
\begin{equation}\label{Dif_Eq_C}
\begin{split}
      \frac{\partial f_a}{\partial x}=-4&f_a|P(k)|^2-2f_bP^{*}(k), \\
      \frac{\partial f_b}{\partial x}&=2f_aP(k).    
\end{split}
\end{equation}
Substitute the second equation to the first equation, one has
\begin{equation}\label{PDE}
	\frac{\partial^2 f_a}{\partial x^2}+4|P(k)|^2\frac{\partial f_a}{\partial x}+4|P(k)|^2 f_a=0.
\end{equation}
Notice that such equation is a standard second-order partial differential equation. Its solution has been discussed thoroughly. In general, the solution of can be given by
\begin{equation}\label{solution}
\begin{split}
      f_a&=C_1 e^{q_1 x}+C_2 e^{q_2 x},\ \qquad\qquad\qquad \Delta > 0\\
      f_a&=(C_1+C_2 x)e^{q_1 x}, \ \ \qquad\qquad\qquad  \Delta = 0\\     
      f_a&=e^{\gamma x}(C_1\cos{\beta x}+C_2\sin{\beta x}), \qquad\Delta < 0
\end{split}
\end{equation}
where $q_{1,2}=(-4|P(k)|^2\pm\sqrt{\Delta})/2$ with $\Delta=16|P(k)|^4-16|P(k)|^2$. $C_1$ and $C_2$ are constants depending on initial conditions. $\gamma$ and $\beta$ are the real and imaginary part of complex $q_{1,2}$ when $\Delta<0$. In our case, $|P(k)|< 1$ so that $\Delta<0$. The case when $|P(k)|=1$ means that state $|k\rangle$ can be searched with one step, which is trivial and is not considered here. Thus, the solution to Eq. (\ref{PDE}) is
\begin{equation}\label{SOL}
	f_a=e^{-2|P(k)|^2x}(C_1\cos{(2\Tilde{\Delta} x)}+C_2\sin{(2\Tilde{\Delta} x)}),
\end{equation}
where $\Tilde{\Delta}=\sqrt{|P(k)|^2-|P(k)|^4}$. The dynamics given by Eq. (\ref{SOL}) is a damping oscillation. The period of the oscillation is $T=\pi/\Tilde{\Delta}$. It indicates that in the time of $T$, there is a moment when $f_a$ takes its maximum, approaching to be one. Therefore, the steps number for the search is in the order of $T$. The speedup of the Grover search under the condition can be verified by comparing the order of $T$ and $s$. More strictly, one has the condition
\begin{equation}\label{GeneralCondition}
     \max\limits_{k=1,\dots,N}\left\{\Tilde{\Delta}^{-1}(k)\right\}<\min\limits_{j=1,\dots,N}\left\{\frac{1}{p_j}\right\}.
\end{equation}
The condition given by Eq. (\ref{GeneralCondition}) indicates a global speed up over the classical treatment. Notice that we omit the constant factor $\pi$ because it does not affect the order. If one limits the problem to searching the $k$th element in the database, the condition can be loosen to 
\begin{equation}
    \Tilde{\Delta}^{-1}(k)<\frac{1}{p_k}=\frac{1}{|P(k)|^2}.
\end{equation}
Then, because $|P(k)|$ is not zero generally,
\begin{equation}
    |P(k)|<\sqrt{1-|P(k)|}.
\end{equation}
Such a condition indicates a local speed up over the classical treatment, which is only effective for searching one element. Obviously, the inequality holds when $|P(k)|<1/2$. It is easy to satisfy such condition when $N$ is sufficiently large.

In what follows, we provide two specific examples of the above general analysis. In the first example, we show that the original unstructured search by Grover's idea can be obtained from our consideration. In the second, we show the results when the distribution of database is that of a coherent state. We also check the effectiveness of the contentious form given by Eq. (\ref{Dif_Eq_C}) in the two examples. 

\section*{Two examples}\label{sec:Grover}
\begin{figure}[htbp]
\centering
\includegraphics[width=4.0in]{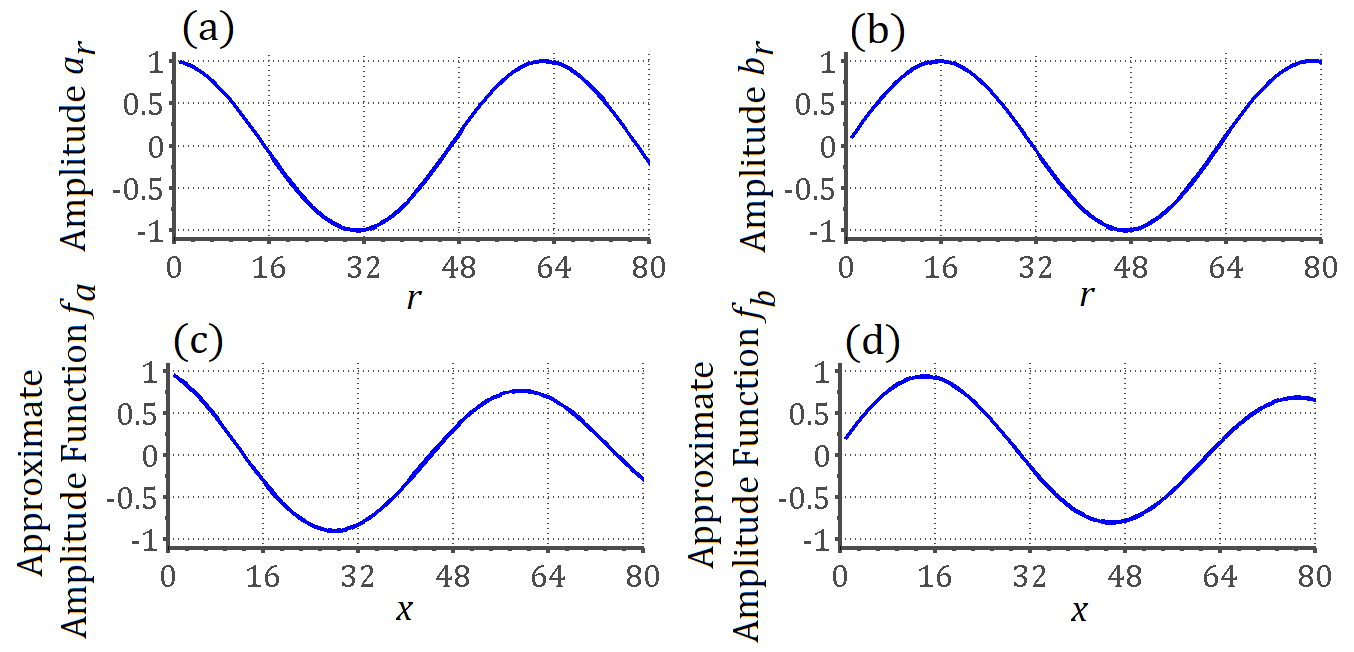}
\caption{The plot of (a) the amplitude $a_r$ and (b) $b_r$ obtained by solving Eq. (\ref{Dif_Eq}) when $P(k)=1/\sqrt{N}$. The plot of the approximation $f_a(x)$ and $f_b(x)$ obtained by solving Eq. (\ref{Dif_Eq_C}) are shown in (c) and (d).}\label{Fig:Check_1}
\end{figure}

{\it{I. Back to unstructured search.}}  The case of unstructured search can be easily obtained by setting $P(k)=1/\sqrt{N}$. The solutions of Eq. (\ref{Dif_Eq}) and the approximation in this example are compared in Fig. \ref{Fig:Check_1}, when $N=20$. The amplitude of $a_r$ and $b_r$ are obtained by solving Eq. (\ref{Dif_Eq}) step-by-step under the condition $a_1=1-4|P(k)|^2$ and $b_1=2P(k)$. The approximation $f_a(x)$ and $f_b(x)$ are obtained by using the 3rd equation of Eqs. (\ref{solution}) under the condition $f_a(0)=1-4|P(k)|^2$ and $f_b(0)=2P(k)$. From the plots, it can be seen that the locations of the maximum of $a_r$ and $b_r$ are nearly the same with those of $f_a$ and $f_b$. Hence, the approximation is valid.

Then, the scale of the step number for the search problem can be estimated by
\begin{equation}
    \Tilde{\Delta}=\sqrt{\frac{1}{N}-\frac{1}{N^2}}=\sqrt{\frac{N-1}{N^2}}.  
\end{equation}
When $N$ is big enough, one has $\sqrt{(N-1)/N^2}\approx 1/\sqrt{N}$. Therefore, the step number for Grover search is in the order of $\sqrt{N}$. It worth mentioning that, in such a case, because $|P(k)|=1/\sqrt{N}\to 0$ for big $N$, the factor $e^{-2|P(k)|x}$ is close to one. It guarantees $f_a$ finally approaches to one. 

The classical search algorithm on the unstructured database is basically checking each elements in the database one by one. Because the probability of finding one element is $1/N$, the step number for the searching by classical treatment is in the order of $N$. Hence, a quadratic speedup can be observed by comparing the order of the two step numbers.
\\
\begin{figure}[htbp]
\centering
\includegraphics[width=3.2in]{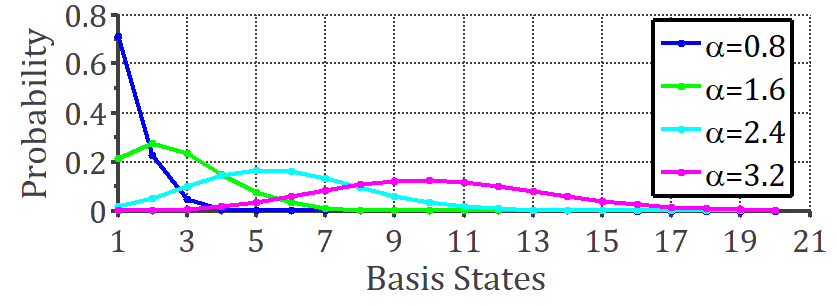}
\caption{The probability distribution of $|\alpha'\rangle$ when $|\alpha|=0.8,\ 1.6,\ 2.4,~\rm{and}\ 3.2$. We take $q_1=1$ and $N=20$. The cases of other $q_1$ and $N$ are similar.}\label{Fig:Prob}
\end{figure}

\begin{figure}[htbp]
\centering
\includegraphics[width=4.1in]{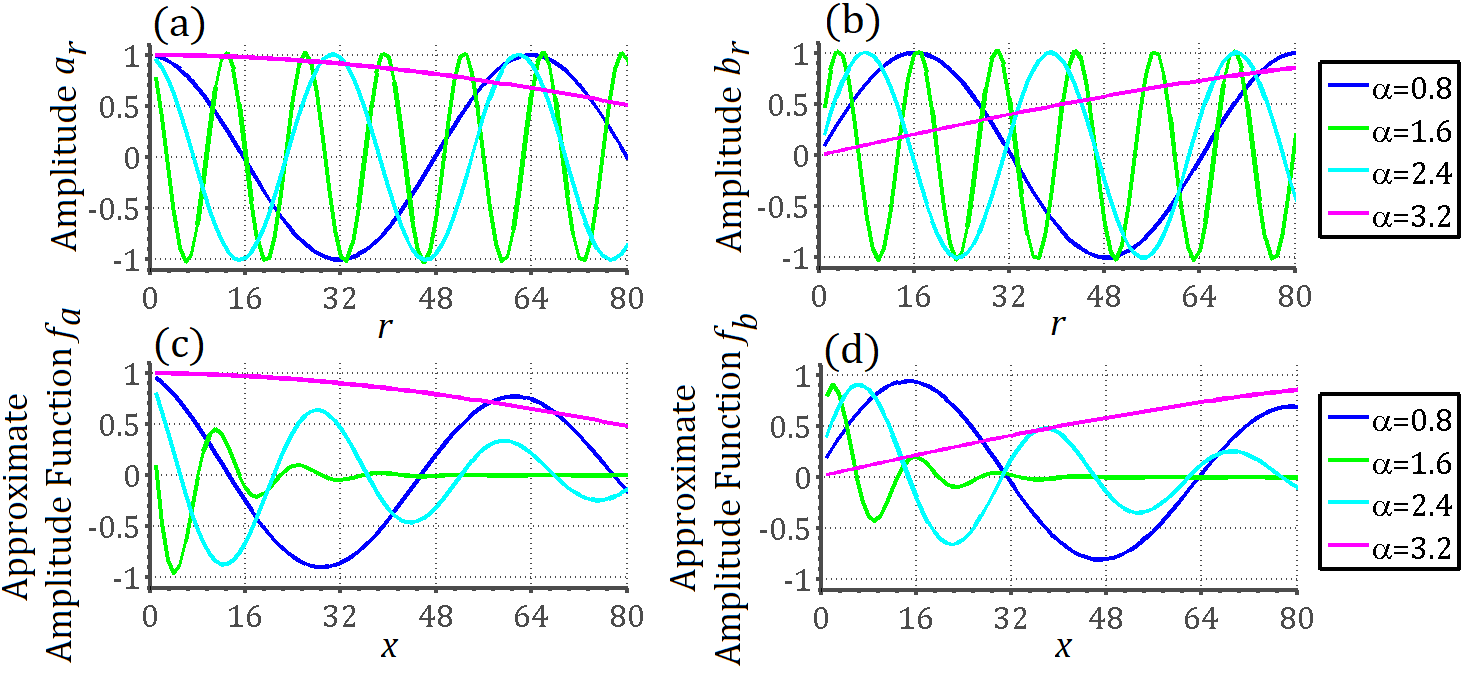}
\caption{The plot of (a) the amplitude $a_r$ and (b) $b_r$ obtained by solving Eq. (\ref{Dif_Eq}) when $P(k)$ is given by Eq. (\ref{eq:coherentPK}). The plot of the approximation $f_a(x)$ and $f_b(x)$ obtained by solving Eq. (\ref{Dif_Eq_C}) are shown in (c) and (d). We set the target state 
 to be the one when $k=3$, and the other parameter setup is the same with the plot in Fig. \ref{Fig:Prob}.}\label{Fig:Check_2}
\end{figure}

\noindent{\em II. Grover search by using coherent state.}
In this part, we consider the case when the distribution of database $\{(y_1,p_1),\dots,(y_N,p_N)\}$ satisfies (or partially satisfies) the distribution of the coherent state. The coherent state in the particle number basis can be expressed by
\begin{equation}\label{CS}
    |\alpha\rangle=e^{-\frac{1}{2}|\alpha|^2}\sum^{\infty}_{q=0}\frac{\alpha^q}{\sqrt{q!}}|q\rangle,
\end{equation}
where $\alpha$ is a complex number. Such a state is a natural state in optical amplification cavity. Notice that there are infinite basis states in the coherent state. Therefore, for finite databases, one can consider encoding them into parts of the state (\ref{CS}). Define the $N$-dimensional database state $|\alpha'\rangle$,

\begin{equation}
    |\alpha'\rangle=N_q\sum^{q_1+N}_{q=q_1}\frac{e^{-\frac{1}{2}|\alpha|^2}\alpha^q}{\sqrt{q!}}|q\rangle,
\end{equation}
where $N_q$ is the normalization factor, given by
\begin{equation}
    N_q=\left[\sum^{q_1+N}_{q=q_1}\frac{e^{-|\alpha|^2}|\alpha|^{2q}}{q!}\right]^{-\frac{1}{2}}.
\end{equation}
Thus, for a target state $|k\rangle$ in the database, one has,
\begin{equation}\label{eq:coherentPK}
    \begin{split}
        |P(k)|&=\frac{e^{-\frac{1}{2}|\alpha|^2}|\alpha|^k}{\sqrt{k!}}\cdot N_q \\
        &=\frac{|\alpha|^k}{\sqrt{k!}}\left[\sum^{q_1+N}_{q=q_1}\frac{|\alpha|^{2q}}{q!}\right]^{-\frac{1}{2}}. \ \ (q_1\le k\le q_1+N)
    \end{split}    
\end{equation}
Notice that, when $q_1$ is large enough, $|\alpha|^k/\sqrt{k!}$ slowly varies with $k$. Thus, the case will go back to the unstructured database, as shown in the first part of this section. 
When $q_1$ is not large enough, the magnitude of $|P(k)|$ is given by $\alpha$ and $q_1$. We numerically provide several cases shown in Fig. \ref{Fig:Prob}. The solutions of Eq. (\ref{Dif_Eq}) and the approximation in this example are compared in Fig. \ref{Fig:Check_2}. The searching target is the basis state when $k=3$, and the rest parameters are set to be the same with those for Fig. \ref{Fig:Prob}. The amplitude of $a_r$ ($b_r$) and its approximation $f_a(x)$ ($f_b(x)$) are obtained by the same method of the first example. The initial conditions are also the same. From the plots, it can be seen that the locations of the maximum of $a_r$ and $b_r$ are also approximately the same with those of $f_a$ and $f_b$. The major difference lies in the damping of $f_a$ and $f_b$, resulted from the factor $e^{-2|P(k)|^2x}$. In the finite cases, one could just neglecting the factor for the estimation of the locations of the maximum. Because such a factor has no affection on the locations. Further, as in the first example, it can be seen that the factor approaches to one when $|P(k)|$ is small enough. Such a condition is better met if the database is large. Therefore, the method can be used to estimate the scale of step number when the size of database goes to infinity. 

\begin{figure}[htbp]
\centering
\includegraphics[width=3.6in]{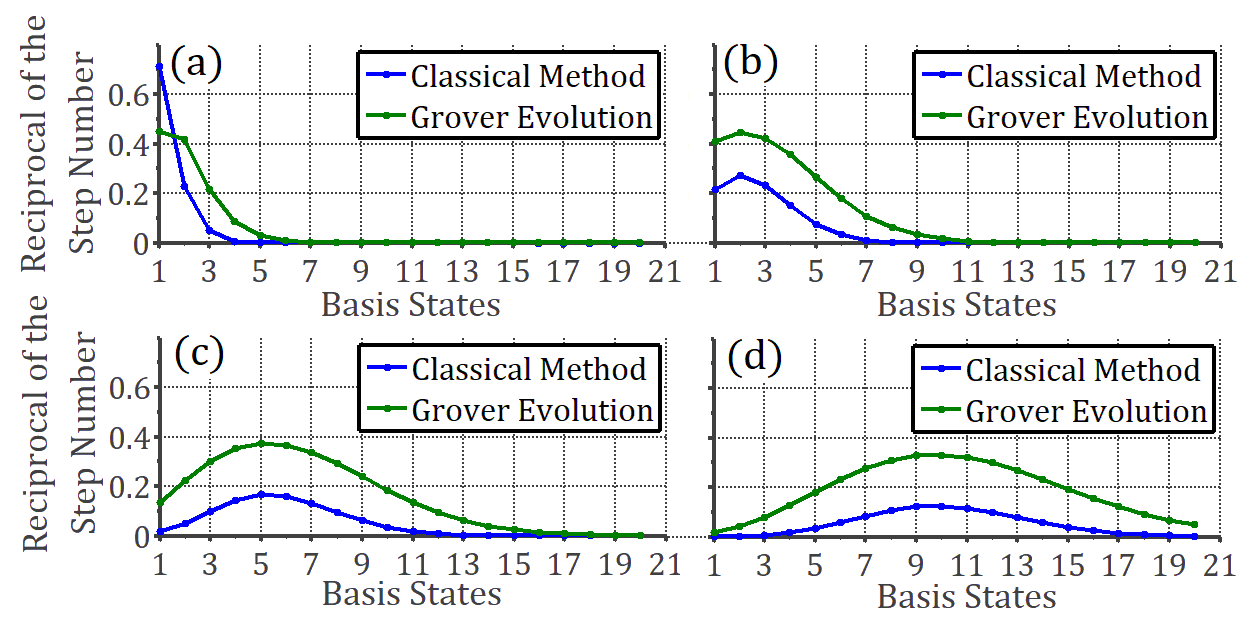}
\caption{The comparison of the reciprocals of step numbers when searching a basis state $|k\rangle$ ($k=1,...,20$) by classical and Grover treatments. The $y$-axis represents the reciprocal of the step numbers. The value of $\alpha$ is 0.8 in (a), 1.6 in (b), 2.4 in (c), and 3.2 in (d).}\label{Fig:C1}
\end{figure}
\begin{figure}[htbp]
\centering
\includegraphics[width=3.6in]{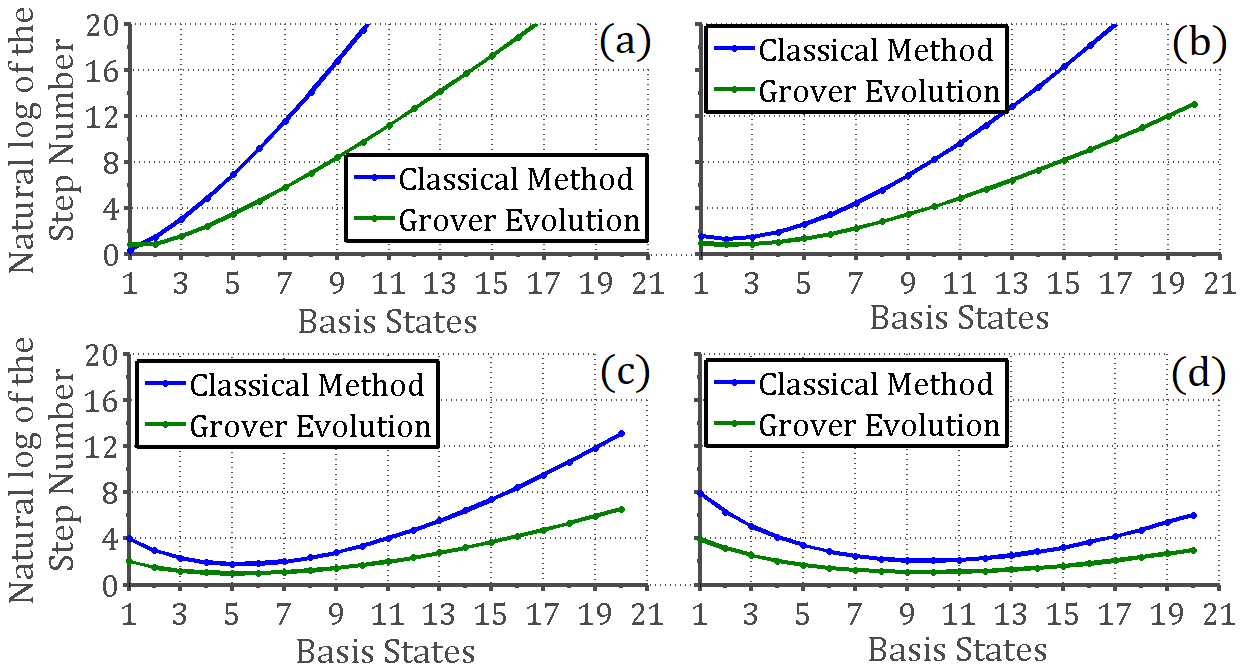}
\caption{The comparison of the natural logarithm of step numbers when searching a basis state $|k\rangle$ ($k=1,...,20$) by classical and Grover treatments. The $y$-axis represents the natural logarithm of the step numbers. The value of $\alpha$ is also 0.8 in (a), 1.6 in (b), 2.4 in (c), and 3.2 in (d).}\label{Fig:C2}
\end{figure}
By substituting the probability distribution to $\Tilde{\Delta}$, the order of the step number can be estimated by the magnitude of $\Tilde{\Delta}$. The step number of the classical treatment is obtained by $1/p_k$ when searching for $y_k$. In order to show a clear comparison, we firstly compare the reciprocal of the step numbers of the two cases, and the results is shown in Fig. \ref{Fig:C1}. In fact, reciprocal of the step number in classical search for $y_k$ is just its probability $p_k$. Extended from the classical concepts, the reciprocal of the step number in Grover search, which is $\Tilde{\Delta}$, represents an effective probability of ``searching for $y_k$''. The comparison between the two reciprocals are equivalent to the comparison of the step number, but with an inverse trend. The smaller step number indicates a larger reciprocal, and vice versa. From the results, we can see that in general, the Grover evolution is a better strategy over the classical treatment. An exception occurs in Fig. \ref{Fig:C1}(a), when searching for the first element. This is because $\alpha$ is relatively small in such a case. We secondly compare the natural logarithm of the step numbers of the cases, and the results are shown in Fig. \ref{Fig:C2}. By Fig. \ref{Fig:C2}, a clear advance in steps number can be observed, and the exception also occurs in (a). The results in Fig. \ref{Fig:C1} and Fig. \ref{Fig:C2} indicate that the Grover search on a  database distributed in the probability given by the coherent state is able to show an advance over the classical methods. According to our conditions in the second section, 
such an advance belongs to the local speedup. 

\section*{Conclusion}
By using the property of differential equation, we investigate the application of Grover's algorithm for weighted database searches, a prevalent scenario in practical settings. 
We calculate the requisite steps for the Grover search and contrast these calculations with classical methodologies. Our analysis differs from the previous method, by considering the asymptotic behaviour of the Grover evolution. This allows us to introduce the tools for analysing differential equations, so that performance of the Grover search on unstructured database can be well assessed. Through detailed analysis, we pinpoint the specific conditions conducive to acceleration through Grover's algorithm. To illustrate our discoveries, we showcase two compelling examples. The first example validates our theoretical framework by aligning closely with Grover's seminal outcomes. In the second example, we demonstrate how implementing Grover's search on a database governed by a probability distribution resembling a coherent state yields significant speed enhancements compared to traditional methods. Meanwhile, we specify that the approximation strategy are effective in the samples, by comparing the results with the explicit solutions of the difference Eqs. (\ref{Dif_Eq}). These results represent a significant advancement of 
analysing Grover algorithm, enriching its implementation strategies and broadening its scope of potential applications.

\section*{Acknowledgements}
This work was funded by MCIN/AEI /10.13039/501100011033 (No. PID2021-126273NB-I00), by ``ERDF A way of making Europe'', and by the Basque Government through Grant No. IT1470-22. We also acknowledge the support by National Natural Science Foundation of China (No.11904022). This project has also received support from the Spanish Ministry for Digital Transformation and of Civil Service of the Spanish Government through the QUANTUM ENIA project call - Quantum Spain, EU through the Recovery, Transformation and Resilience Plan-NextGenerationEU within the framework of the Digital Spain 2026.

\section*{Author contributions statement}
The idea is initialized by L.-A. W. The theoretical derivation is performed by Y. S. All authors reviewed the manuscript. 

\section*{Data availability}
The datasets used and/or analysed during the current study available from the corresponding author on reasonable request.


\begin{thebibliography}{99}
\bibitem{Grover1997}
Grover, L. K. Quantum mechanics helps in searching for a needle in a haystack. {\it \prl{79}}, 325 (1997).

\bibitem{Feynman1982}
Feynman, R. P. Simulating physics with computers. {\it Int. J. Theor. Phys.} {\bf 21}, 467 (1982).

\bibitem{Karp1972}
Karp, R. M. Reducibility among combinatorial problems, in
\emph{Complexity of Computer Computations} (Springer, Berlin, 1972),
pp. 85–103.

\bibitem{Weibe2015}
Wiebe, N., Kapoor, A., \& Svore, K. Quantum perceptron models, arXiv:1602.04799.

\bibitem{Gilliam2021}
Gilliam, A.,  Woerner, S., \& Gonciulea, C. Grover adaptive search for constrained polynomial binary optimization. {\it Quantum} {\bf 5}, 428 (2021).

\bibitem{Brassard2002}
Brassard, G., Hoyer, P., Mosca, M., \&  Tapp, A. Quantum amplitude amplification and estimation. {\it AMS Contemp. Math.} {\bf 305}, 53 (2002).

\bibitem{Ren2020}
Ren, F.-H., Wang, Z.-M., \& Wu, L.-A. Accelerated adiabatic quantum search algorithm via pulse control in a non-Markovian environment. {\it \pra{102}}, 062603 (2020).

\bibitem{Wu2002}
Wu, L.-A., Byrd, M. S., \& Lidar, D. A. Polynomial-time simulation of pairing models on a quantum computer. {\it \prl{89}}, 057904 (2002). 

\bibitem{Pyshkin2016}
Pyshkin, P. V., Luo, D.-W., Jing, J., You, J. Q., \& Wu, L.-A. Expedited holonomic quantum computation via net zero-energy-cost control in decoherence-free subspace. {\it Sci. Rep.} {\bf 6}, 37781 (2016).

\bibitem{Sun2020}
Sun, Y., Zhang, J.-Y., Byrd, M. S., \& Wu, L.-A. Trotterized adiabatic quantum simulation and its application to a simple all-optical system. {\it New J. Phys.} {\bf 22} 053012 (2020).

\bibitem{Tezuka2022}
Tezuka, H., Nakaji, K., Satoh, T., \& Yamamoto, N. Grover search revisited: Application to image pattern matching. {\it \pra{105}}, 032440 (2022).

\bibitem{Reitzner2019}
Reitzner, D. \&  Hillery, M. Grover search under localized dephasing. {\it \pra{99}}, 012339 (2019).

\bibitem{Brickman2005}
Brickman, K.-A., Haljan, P. C., Lee, P. J., Acton, M.,  Deslauriers, L., \& Monroe, C. Implementation of Grover’s quantum search algorithm in a scalable system. {\it \pra{72}}, 050306(R) (2005).

\bibitem{Figgatt2017}
Figgatt, C. Maslov, D., Landsman, K. A., Linke, N. M.,  Debnath, S., \&  Monroe, C. Complete 3-qubit Grover search on a programmable quantum computer. {\it Nat. Commun.} {\bf 8}, 1918 (2017).

\bibitem{Vemula2205}
Vemula, D. R., Konar, D.,  Satheesan, D., Kalidasu, S. M., \& Cangi, A. A scalable 5,6-qubit Grover's quantum search algorithm, arXiv: 2205.00117.

\bibitem{Gustiani2021}
Gustiani, C. \& DiVincenzo, D. P. Blind three-qubit exact Grover search on a nitrogen-vacancy-center platform. {\it 
 \pra{104}}, 062422 (2021).

\bibitem{Pan2021}
Pan, N., Chen, T., Sun, H. \& Zhang, X. Electric-Circuit Realization of Fast Quantum Search. {\it Research} {\bf 2021}, 9793071 (2021).

\bibitem{Ji2022}
Ji, T., Pan, N., Chen, T., \& Zhang, X. Fast quantum search of multiple vertices based on electric circuits. {\it Quantum Inf. Proc.}, {\bf 21}, 172 (2022).

\bibitem{Ji2022-2}
Ji, T., Pan, N., Chen, T. \&  Zhang, X. Quantum search of many vertices on the joined complete graph. {\it Chin. Phys. B.} {\bf 31}, 070504 (2022).

\bibitem{Biham1999}
Biham, E., Biham, O., Biron, D., Grass, M., \& Lidar, D. A. Grover’s quantum search algorithm for an arbitrary initial amplitude distribution. \pra{60}, 2742 (1999).

\bibitem{Biham2000}
Biham, E., Biham, O., Biron, D., Grass, M. Lidar, D. A., \& Shapira, D.  Analysis of generalized Grover quantum search algorithms using recursion equations. \pra{63}, 012310 (2000).

\bibitem{Nielsen2000}
Nielsen, M. A. \& Chuang, I. L. {\it Quantum computation and quantum information}, Cambridge University Press, Cambridge (2000).
\end{thebibliography}
\end{document}